\documentclass[a4paper,10pt,conference]{IEEEtran}
\usepackage[compress]{cite}
\usepackage[utf8]{inputenc}
\usepackage{amsmath}
\usepackage{amssymb}
\usepackage{mathrsfs}
\usepackage{enumerate}
\usepackage{tikz}
\usepackage{adjustbox}
\usepackage{xcolor}
\usepackage{caption}
\usetikzlibrary{shapes,arrows,chains}
\usepackage{subfig}
\usepackage{graphicx}
\usepackage{enumerate}
\usepackage{algorithm}
\usepackage{algpseudocode}
\usepackage{xpatch}
\usepackage{mathtools}
\usepackage{resizegather}
\DeclarePairedDelimiter{\norm}{\lVert}{\rVert}
\algdef{SE}[DOWHILE]{Do}{doWhile}{\algorithmicdo}[1]{\algorithmicwhile\ #1}%
\algrenewcommand\alglinenumber[1]{{\sffamily\footnotesize#1}}
\makeatletter
\xpatchcmd{\algorithmic}{\itemsep\z@}{\itemsep=.25ex plus2pt}{}{}
\makeatother

\usepackage{color}
\usepackage{soul}
\usepackage{environ}         
\usepackage{etoolbox}        
\usepackage{graphicx}        
\usepackage[acronym]{glossaries}
\newacronym{1g}{1G}{first-generation}
\newacronym{4g}{4G}{fourth-generation}
\newacronym{5g}{5G}{fifth-generation}
\newacronym{mimo}{MIMO}{multiple-input-multiple-output}
\newacronym{siso}{SISO}{single-input-single-output}
\newacronym{mamimo}{MaMIMO}{massive multiple-input-multiple-output}
\newacronym{sumimo}{SU-MIMO}{single user MIMO}
\newacronym{mumimo}{MU-MIMO}{multi user MIMO}
\newacronym{embms}{eMBMS}{evolved Multimedia Broadcast and Multicast Service}
\newacronym{sca}{SCA}{successive convex approximation}
\newacronym{sinr}{SINR}{signal-to-interference-plus-noise ratio}
\newacronym{ula}{ULA}{uniform linear array}
\newacronym{mcs}{MCS}{modulation and coding scheme}
\newacronym{mrt}{MRT}{maximum ratio transmission}
\newacronym{zf}{ZF}{zero-forcing}
\newacronym{se}{SE}{spectral efficiency}
\newacronym{ase}{ASE}{aggregated SE}
\newacronym{asd}{ASD}{angular standard deviation}
\newacronym{adr}{ADR}{aggregated data rate}
\newacronym{embb}{eMBB}{enhanced mobile broadband}
\newacronym{mmtc}{mMTC}{massive machine type communications}
\newacronym{urllc}{URLLC}{ultra reliable low latency communications}
\newacronym{csi}{CSI}{channel state information}
\newacronym{cqi}{CQI}{channel quality indicator}
\newacronym{pmi}{PMI}{precoding matrix indicator}
\newacronym{ri}{RI}{rank indicator}
\newacronym{csi-rs}{CSI-RS}{CSI-reference signal}
\newacronym{cri}{CRI}{CSI-RS resource indicator}
\newacronym{bs}{BS}{base station}
\newacronym{re}{RE}{resource element}
\newacronym{mmwave}{mmWave}{millimeter-wave}
\newacronym{umwave}{$\mu$mWaves}{micrometer waves}
\newacronym{rnn}{RNN}{recurrent neural network}
\newacronym{cnn}{CNN}{convolutional neural network}
\newacronym{ngmn}{NGMN}{next-generation mobile network}
\newacronym{lte}{LTE}{Long Term Evolution}
\newacronym{lte-a}{LTE-A}{Long Term Evolution Advanced}
\newacronym{5gnr}{5G NR}{5G New Radio}
\newacronym{phy}{PHY}{physical}
\newacronym{mac}{MAC}{medium access control}
\newacronym{3gpp}{3GPP}{3rd Generation Partnership Project}
\newacronym{fdd}{FDD}{frequency division duplexing}
\newacronym{tdd}{TDD}{time division duplexing}
\newacronym{ofdm}{OFDM}{orthogonal frequency division multiplexing}
\newacronym{ss}{SS}{synchronization signal} 
\newacronym{pss}{PSS}{primary synchronization signal} 
\newacronym{sss}{SSS}{secondary synchronization signal} 
\newacronym{pbch}{PBCH}{physical broadcast channel} 
\newacronym{dmrs}{DMRS}{demodulation reference signal} 
\newacronym{gnb}{gNB}{next generation nodeB} 
\newacronym{rsrp}{RSRP}{reference signal received power} 
\newacronym{rrm}{RRM}{radio resource management} 
\newacronym{srs}{SRS}{sounding reference signal} 
\newacronym{ran}{RAN}{radio access network} 
\newacronym{nn}{NN}{neural network} 
\newacronym{ue}{UE}{user equipment} 
\newacronym{awgn}{AWGN}{additive white Gaussian noise} 
\newacronym{epa}{EPA}{Extended Pedestrian A model}
\newacronym{eva}{EVA}{Extended Vehicular A model}
\newacronym{etu}{ETU}{Extended Typical Urban model}
\newacronym{tdl}{TDL}{tapped delay line}
\newacronym{cdl}{CDL}{clustered delay line}
\newacronym{uma}{UMa}{urban macro-cell}
\newacronym{isd}{ISD}{inter-site distance}
\newacronym{nlos}{NLOS}{non-line of sight}
\newacronym{los}{LOS}{line of sight}
\newacronym{o2o}{O2O}{outdoor-to-outdoor}
\newacronym{o2i}{O2I}{outdoor-to-indoor}
\newacronym{ul}{UL}{uplink}
\newacronym{dl}{DL}{downlink}
\newacronym{ls}{LS}{least squares}
\newacronym{mmse}{MMSE}{minimum mean square error}
\newacronym{snr}{SNR}{signal-to-noise ratio}
\newacronym{mse}{MSE}{mean square error}
\newacronym{nr}{NR}{New Radio}
\newacronym{prb}{PRB}{physical resource block}
\newacronym{scs}{SCS}{subcarrier spacing}
\newacronym{bler}{BLER}{block error rate}
\newacronym{smmmra}{SMMMRA}{subgroup multicast \gls{mamimo} resource allocation}
\newacronym{mmf}{MMF}{max-min fairness}
\newacronym{smmu}{SMMU}{subgroups of multicast \gls{mamimo} users}
\newacronym{gsmma}{GSMMA}{greedy subgroup multicast \gls{mamimo} algorithm}

\include{cite}
\usepackage{soul}
\usepackage[normalem]{ulem}

\begin{document}
\title{User Subgrouping in Multicast Massive MIMO over Spatially Correlated Rayleigh Fading Channels}
\author{\IEEEauthorblockN{Alejandro de la Fuente\IEEEauthorrefmark{1}\IEEEauthorrefmark{2}, Giovanni Interdonato\IEEEauthorrefmark{3}, Giuseppe Araniti\IEEEauthorrefmark{2}}
\IEEEauthorblockA{\IEEEauthorrefmark{1}Department of Signal Theory and Communications, University Rey Juan Carlos}
\IEEEauthorblockA{\IEEEauthorrefmark{2}DIIES Department, University Mediterranea of Reggio Calabria}
\IEEEauthorblockA{\IEEEauthorrefmark{3}Department of Electrical and Information Engineering, 
University of Cassino and Southern Lazio}
Email: alejandro.fuente@urjc.es}

\maketitle
\begin{abstract}
\Gls{mamimo} multicasting has received significant attention over the last years. \Gls{mamimo} is a key enabler of 5G systems to achieve the extremely demanding data rates of upcoming services. Multicast in the physical layer is an efficient way of serving multiple users, simultaneously demanding the same service and sharing radio resources. 
This work proposes a subgrouping strategy of multicast users based on their spatial channel characteristics 
to improve the channel estimation and precoding processes. We employ \gls{mmf} power allocation strategy to maximize the minimum \gls{se} of the multicast service. Additionally, we explore the combination of spatial multiplexing with orthogonal (time/frequency) multiple access. By varying the number of antennas at the \gls{bs} and users' spatial distribution, we also provide the optimal subgroup configuration that maximizes the spectral efficiency per subgroup. Finally, we show that serving the multicast users into two orthogonal time/frequency intervals offers better performance than only relying on spatial multiplexing. 
\end{abstract}

\glsresetall
\section{Introduction}

Upcoming mobile services demand stringent performance requirements, both in terms of data rates, latency, and the number of connected devices \cite{2019Ericsson}. \Gls{mamimo} plays a key role in \gls{5g} systems to fulfill these requirements \cite{2014Andrews,2014Boccardi}. This technology makes use of many antennas at the \gls{bs} to jointly and coherently serve multiple users in the same time/frequency resources by yielding array gain, spatial multiplexing gain, and spatial diversity \cite{2010Marzetta,2014Larsson}.
Importantly, \gls{mamimo} can offer, in most of the propagation environments, two fundamental features that are known as \emph{favorable propagation} and \emph{channel hardening}: as the number of BS antennas increases, users' channels become nearly pairwise orthogonal and deterministic, respectively \cite{Bjornson2016}.
All this leads to a significant increase in spectral and energy efficiency.


\Gls{lte-a} systems fully support broadcast/multicast transmissions through the use of the \gls{embms} \cite{2016delaFuente,2017Araniti}. The \gls{embms} is implemented as an \gls{lte-a} subsystem to share the physical resources between unicast and multicast transmissions. The standard allows the system an efficient resource utilization when multiple users simultaneously demand the same content.

We recently witnessed an increasing interest in multicast \gls{mamimo} transmissions in \gls{mmwave} and sub-6 GHz bands. 
The authors in \cite{2013Yang} propose a multicast \gls{mamimo} strategy using a unique pilot sequence for all the multicast users receiving the same service. This pilot sequence is used in a power control scheme to equalize the throughput among all the users. Sadeghi et al. have extended this proposal to multi-group multicast joint with unicast services in multicell deployments. They have also developed low complexity solutions for multicast and unicast precoders \cite{2017SadeghiTWC,2018SadeghiTWC1,2018SadeghiTWC2}. In \cite{2019Dong}, the authors present a framework to achieve optimal multicast beamforming. The low-dimensional structure in the optimal solution benefits the numerical computation in large antenna systems. In \cite{2019Biason}, the authors show how to shape the beams to deliver multicast information to the users in \gls{mmwave}. They demonstrate that restricting the wireless links to be unicast only may be strongly suboptimal. Furthermore, the authors in \cite{2020Samuylov} develop a mathematical framework to estimate the parameters of the \gls{mmwave} \glspl{bs} for handling a mixture of multicast and unicast sessions. This framework allows the network designers to achieve a lower bound on the required density of the \glspl{bs}.

To the best of our knowledge, existing works on \gls{mamimo} multicasting employ uncorrelated fading channel models. 
However, practical scenarios must take into account spatial correlations which affect the \emph{favorable propagation}, 
mainly in a \emph{poor-scattering} propagation environment. This effect reduces mutual-orthogonality among different users. 

\textbf{Contributions:} We consider spatially correlated fading channels to study the delivery of multicast service. We propose a user subgrouping method based on the level of the user mutual-orthogonality. Our strategy reduces the pilot contamination making precoding more effective. We also consider optimal \gls{mmf} power control and optimize the number of subgroups that maximizes the \gls{se}. Finally, we analyze the benefits of splitting the multicast users into two different time/frequency schedules.

\section{System Model}
\label{sec:system_model}
Let us consider a multicast transmission in a single-cell \gls{mamimo} system with fully digital precoding. We assume the system has a \gls{bs} equipped with a \gls{ula} with $M$ antennas serving  $K$ single-antenna users. We denote the set of multicast users as $\mathcal{K}$, i.e., $\mathcal{K} = \{1,\ldots, K\}$. Let $\boldsymbol{h}_{gk} \in \mathbb{C}^{M \times 1}$ be the channel between the \gls{bs} and user $k$ included in subgroup $g$ (we detail the subgrouping model in subsection \ref{sec:subgrouping}). The block-fading model is herein assumed.

Although uncorrelated Rayleigh fading channels are widely employed to study \gls{mamimo} multicasting performance \cite{2017SadeghiTWC}, it is more likely to have spatially correlated fading in practical scenarios. 
%
Hence, 
\begin{equation}
    \begin{split}
        \boldsymbol{h}_{gk}  \sim \mathcal{CN}(\boldsymbol{0},\boldsymbol{R}_{gk}),
    \end{split}
\end{equation}  
where $\boldsymbol{R}_{gk} \in \mathbb{C}^{M \times M}$ is the positive semi-definite spatial correlation matrix at user $k$ in subgroup $g$, incorporating path-loss, shadowing, and spatial correlation fading. 
We denote the corresponding large-scale fading coefficient as 
\begin{equation}
    \begin{split}
        {\beta}_{gk}  = \frac{1}{M} \text{tr}\left(\boldsymbol{R}_{gk}\right),
    \end{split}
    \label{eq:beta}
\end{equation}  
while the small-scale fading follows a Rayleigh distribution.



We parameterize the correlation matrices $\boldsymbol{R}_{gk}$ using the azimuth angles from the \gls{bs} to the users. 
%
%
The \gls{bs} receives from user $k$ a signal that consists of the superposition of $N$ multipath components. Each multipath component reaches the BS as a planar wave from a particular angle $\varphi_k(n) \in \left[\Phi_k,\Phi_k + \phi_k\right]$ for $n=1,\ldots,N$, where $\Phi_k$ is the 2D nominal angle between user $k$ and the \gls{bs} and $\phi_k$ is a random deviation from the nominal angle whose standard deviation in radians is called \gls{asd}. Hence, 
\begin{equation}
    \begin{split}
        \boldsymbol{h}_{gk} = \sum_{n=1}^N \rho_k(n)\boldsymbol{a}_k\left(\varphi_k(n)\right),
    \end{split}
\end{equation}  
where $\rho_k(n) \in \mathbb{C}$ represents the gain and phase of the $n$th physical path for user $k$ and $\boldsymbol{a}_k\left(\varphi_k(n)\right) \in \mathbb{C}^{M}$ is the steering vector of the \gls{ula} given by
\begin{equation}
    \begin{split}
        \boldsymbol{a}_k\left(\varphi_k\right) = \left[1 \ \ e^{j2\pi\delta\cos{\varphi_k}} \ \ \ldots \ \ e^{j2\pi\delta(M-1)\cos{\varphi_k}}\right]^\intercal,   
    \end{split}
\end{equation}  
where $\delta$ is the distance between adjacent antennas, normalized by the wavelength.
Both the nominal angle $\Phi_k$ and the \gls{asd} characterize the spatially correlated Rayleigh fading channels. The \gls{bs} estimates the spatial correlation matrix of each user on the large-scale fading time-scale (i.e., over several coherence blocks). Therefore, we can reasonably assume $\boldsymbol{R}_{gk}, \forall k \in \mathcal{K}$, to be known at the \gls{bs}.

\subsection{Multicast \gls{mamimo} subgrouping}
\label{sec:subgrouping}
Multicast subgrouping consists in splitting the $K$ users, demanding the same multicast service, into $G$ disjoint subgroups based on their channel similarities. The objective of subgrouping the users is to increase the \gls{ase}. A specific and unique transmission setting characterizes each multicast subgroup. Let $\mathcal{K}_g$ and $K_g$ denote the set and the number of multicast users in subgroup $g$, respectively (i.e., $|\mathcal{K}_g|=K_g$). Moreover, let $\mathcal{G}$ be the set of subgroups. Hence, we have
$K = \sum_{g=1}^G K_g$. 
     
\begin{figure}[t]
\centering
\includegraphics[width=0.83\columnwidth]{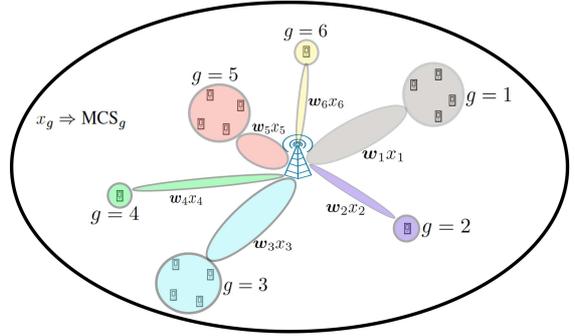}
\caption{Massive MIMO multicasting scenario with user subgrouping based on the spatial characteristics.}
\label{fig:subgroupingscenario}
\end{figure}

Fig. \ref{fig:subgroupingscenario} illustrates an example of user subgrouping-based multicast transmission in a single-cell \gls{mamimo} system. The users are grouped based on the level of orthogonality of their spatial correlation matrices, namely users with similar spatial characteristics belong to the same subgroup.


\subsection{Channel estimation}

We assume that the users in the same subgroup are assigned the same pilot sequence, while mutually orthogonal pilots are assigned over different subgroups. As co-pilot users have linearly dependent channel estimates, the BS cannot separate the users of the same subgroup in the spatial domain.
Consequently, the BS can effectively construct as many precoding vectors as the number of orthogonal pilots (which must be at least equal to $G$), and the same precoding vector is employed to all the users of the same subgroup.

Let $\boldsymbol{\Psi} = \left[\boldsymbol{\psi}_1,\ldots,\boldsymbol{\psi}_G\right] \in \mathbb{C}^{\tau_p \times G}$ be the pilot matrix where $\boldsymbol{\psi}_g$ is the pilot sequence of length $\tau_p$ symbols assigned to all the users in subgroup $g$. 
Without loss of generality, we set $\tau_p = G$ to obtain $G$ mutually orthogonal uplink pilots, $\boldsymbol{\Psi}^{\text{H}}\boldsymbol{\Psi}=\tau_p\boldsymbol{I}_G$ ($G$ is known at the channel estimation stage). The \gls{ul} pilot signal received at the BS is
\begin{gather}
    \boldsymbol{Y} = \sum_{g \in \mathcal{G}}\sum_{k \in \mathcal{K}_g} \sqrt{q_{gk}}\boldsymbol{h}_{gk}\boldsymbol{\psi}^\intercal_g + \boldsymbol{N},
\end{gather}
where $q_{gk}$ is proportional to the \gls{ul} pilot power per user $k \in \mathcal{K}_g$, and $\boldsymbol{N} \in \mathbb{C}^{M \times \tau_p}$ is the \gls{awgn} with i.i.d. elements distributed as $\mathcal{CN}(0,\sigma^2)$. To estimate the channel of user $k$ in subgroup $g$, the \gls{bs} first correlates the received signal with $\boldsymbol{\psi}_g^{*}$, obtaining
\begin{gather}
    \boldsymbol{y}_{gk}^{\text{UL}} = \sqrt{q_{gk}}\tau_p\boldsymbol{h}_{gk} + \sum_{k' \in \mathcal{K}_g \setminus \{k\}} \sqrt{q_{gk'}}\tau_p\boldsymbol{h}_{gk'} + \boldsymbol{n}_k,
\end{gather}
where $\boldsymbol{n}_k \sim \mathcal{CN}(\boldsymbol{0},\sigma^2\boldsymbol{I}_M)$ is \gls{awgn}. Then, the \gls{mmse} channel estimate is \cite[Sec. 3.2]{2017Bjornsonbook}
\begin{gather}
    \hat{\boldsymbol{h}}_{gk} = \sqrt{q_{gk}}\boldsymbol{R}_{gk}\left[\sum_{k' \in \mathcal{K}_g}\left(\tau_p q_{gk'} \boldsymbol{R}_{gk'} + \sigma^2\boldsymbol{I}_{M}\right)\right]^{-1}\boldsymbol{y}_{gk}^{\text{UL}}.
\end{gather}
%
Let $\boldsymbol{h}_g$ denote the composite channel of subgroup $g$ given by 
\begin{gather}
\boldsymbol{h}_g = \tau_p\displaystyle\sum_{k=1}^{K_g}\left(\sqrt{ q_{gk}}\boldsymbol{h}_{gk}\right).
\end{gather}
The \gls{mmse} estimate of the composite channel is
\begin{gather}
    \hat{\boldsymbol{h}}_{g}\!=\!\tau_p\!\!\sum_{k \in \mathcal{K}_g} q_{gk} \boldsymbol{R}_{gk}\!\!\left[\sum_{k \in \mathcal{K}_g}\!\left(\tau_p q_{gk} \boldsymbol{R}_{gk} + \sigma^2\boldsymbol{I}_{M}\right)\right]^{-1}\!\!\!\!\!\boldsymbol{y}_{gk}^{\text{UL}}.
\end{gather}
We stack the $G$ composite channel vectors in a matrix $\hat{\boldsymbol{C}} = [\hat{\boldsymbol{h}}_1,\ldots,\hat{\boldsymbol{h}}_G] \in \mathbb{C}^{M \times G}$ and use $\hat{\boldsymbol{C}}$ to formulate the \gls{zf} precoding vector intended for subgroup $g$ as
\begin{equation}
    \begin{split}
\boldsymbol{w_g} = \frac{\boldsymbol{v}_{g}}{\norm{\boldsymbol{v}_{g}}_2},    \end{split}
    \label{eq:ZF}
\end{equation}
where $\boldsymbol{w}_g \in \mathbb{C}^{M \times 1}$,  $\mathbb{E}\left[\norm{\boldsymbol{w}_g}^2\right]=1$, and $\boldsymbol{v}_{g}$ is the $g$-th column vector of the matrix $\boldsymbol{V}=\hat{\boldsymbol{C}}(\hat{\boldsymbol{C}}^{\text{H}}\hat{\boldsymbol{C}})^{-1}$.

\subsection{Downlink data transmission and spectral efficiency}

We assume that the \gls{bs} transmits data to the multicast users by using \gls{zf} precoding. Specifically, the same precoding vector, \gls{mcs} are employed for all the users in the same subgroup.
Let us denote the data symbols for every user $k \in \mathcal{K}_g$ as $x_g$ (unit variance random variables and uncorrelated), 
with $p_g$ being the transmit power allocated to the multicast subgroup $g$. We assume that the users have access only to the statistical \gls{csi}, i.e., $\mathbb{E}[\boldsymbol{h}^{\text{H}}_{gk}\boldsymbol{w}_g]$. Hence, the \gls{dl} data signal received at user $k \in \mathcal{K}_g$ can be written as
\begin{align}
        y_{gk} &= \sqrt{p_g} \mathbb{E}\left[\boldsymbol{h}^{\text{H}}_{gk}\boldsymbol{w}_g\right] x_g + \sqrt{p_g}\left(\boldsymbol{h}^{\text{H}}_{gk}\boldsymbol{w}_g - \mathbb{E}\left[\boldsymbol{h}^{\text{H}}_{gk}\boldsymbol{w}_g\right]\right) x_g \nonumber  \\
        &\quad + \sum_{g' \in \mathcal{G} \setminus \{g\}} \sqrt{p_{g'}} \boldsymbol{h}^{\text{H}}_{gk}\boldsymbol{w}_{g'}x_{g'} + n_k,
    \label{eq:y_gk}
\end{align}    
where the first term denotes the desired signal, the second term is interference due to the user's lack of \gls{csi}, the third term denotes the inter-subgroup interference and finally $n_k \sim \mathcal{CN}(0,\sigma_k^2)$ is the \gls{awgn}. By invoking the capacity-bounding technique in \cite[Sec. 2.3.4]{2016Marzetta}, which treats the second, third and fourth term of \eqref{eq:y_gk}
as effective uncorrelated noise, a downlink achievable spectral efficiency is given by 
\begin{equation}
    \begin{split}
           \text{SE}_{gk} = \left(1 - \frac{\tau_p}{\tau_c}\right) \text{log}_2\left(1 + \gamma_{gk}\right),
     \end{split}
    \label{eq:SE_gk}
\end{equation}
where $\tau_c$ is the coherence block length and $\gamma_{gk}$ is the effective \gls{sinr} given by
\begin{equation}
    \begin{split}
        \gamma_{gk} = &\frac{p_g\Big|{\mathbb{E}\left[\boldsymbol{h}^{\text{H}}_{gk}\boldsymbol{w}_g\right]}\Big|^2}{\displaystyle\sum_{g'=1}^G p_{g'}\mathbb{E}\left[\Big|{\boldsymbol{h}^{\text{H}}_{gk}\boldsymbol{w}_{g'}}\Big|^2\right] - p_g\Big|{\mathbb{E}\left[\boldsymbol{h}^{\text{H}}_{gk}\boldsymbol{w}_g\right]}\Big|^2 + \sigma_k^2},
     \end{split}
    \label{eq:SINR_gk}
\end{equation}
where the expectations are w.r.t. the channel realizations.

Recall that each subgroup receives the same service but with a different \gls{mcs}, which must support the user's \gls{se} with the worst channel condition in the subgroup. Thus, all the users $k \in \mathcal{K}_g$ experience the \gls{se} given by
\begin{equation}
    \begin{split}
           \text{SE}_{g} = \underset{k \in \mathcal{K}_g}{\text{min}} \text{SE}_{gk}.
     \end{split}
    \label{eq:SE_g}
\end{equation}

\section{MaMIMO multicasting}
\label{sec:SSMMU}


The literature of \Gls{mamimo} multicasting  \cite{2013Yang,2017SadeghiTWC,2018SadeghiTWC1,2018SadeghiTWC2,2019Dong} essentially presents two fundamental delivery strategies. The first option consists of serving each user individually by a unicast transmission as in conventional \gls{mamimo}. This approach limits the number of unicast users to the number of orthogonal pilots (assuming no pilot reuse) and requires the number of BS antennas to be larger than the number of users (for the ZF precoding to be performed). Alternatively, multicast transmission may take place over one joint transmission to the whole multicast group. In this case, only one pilot sequence, \gls{mcs} are used for all the users.

\subsection{User subgrouping based on spatial characteristics}

We propose an alternative strategy to deliver a multicast service in a \gls{mamimo} system, which  consists of forming disjoint subgroups of multicast users, each served with a specific \gls{mcs}. The soundness of this approach was shown in \gls{siso} and \gls{mimo} systems, using both wideband and subband channel information \cite{2013Araniti,2018delaFuente}. 

The proposed subgrouping criterion relies on the spatial characteristics of the users: users with similar spatial characteristics, and thereby which cause much interference to each other, are grouped all together.  
To this end, we consider the normalized channels $h_{gk}/\sqrt{\mathbb{E}\{\norm{\boldsymbol{h}_{gk}}^2\}}$ and $h_{g'k'}/\sqrt{\mathbb{E}\{\norm{\boldsymbol{h}_{g'k'}}^2\}}$ of any pair of multicast users $k \in \mathcal{K}_g$, $k' \in \mathcal{K}_{g'}$. The level of orthogonality of the channel directions, quantified by the  variance of the inner product of the normalized channels
\begin{equation}
    \begin{split}
    \mathbb{V}\left[\frac{\boldsymbol{h}_{gk}^{\text{H}} \boldsymbol{h}_{g'k'}}{\sqrt{\mathbb{E}[\norm{\boldsymbol{h}_{gk}}^2]\mathbb{E}[\norm{\boldsymbol{h}_{g'k'}}^2]}}\right] = \frac{\text{tr}\left(\boldsymbol{R}_{gk} \boldsymbol{R}_{g'k'}\right)}{M^2 \beta_{gk} \beta_{g'k'}},
    \end{split}
    \label{eq:orthogonality-level}
\end{equation}
gives a measure of how much interference the users cause to each other: the larger the value, the smaller the orthogonality level of the channel directions and the higher the mutual interference between the two users.\footnote{$\mathbb{V}$ in \eqref{eq:orthogonality-level} denotes the variance operator. A necessary but not sufficient condition for favorable propagation is that \eqref{eq:orthogonality-level} $\to 0$, as $M\to\infty$ \cite{2017Bjornsonbook}.}


Before pilot assignment and channel estimation, assuming perfect knowledge of the large-scale fading parameters, the \gls{bs} forms the user subgroups such that the users in the same subgroup present similar channel characteristics, hence low levels of orthogonality, i.e., those users for which \eqref{eq:orthogonality-level} returns a large value.




The K-means algorithm and its multiple variants provide a simple method to efficiently cluster the multicast users into disjoint subgroups \cite{2010Jain}. This algorithm aims at finding a partition of the $K$ users into $G$ subgroups, minimizing the \gls{mse} according to the selected metric \cite{2018Riera}. 

\subsection{Max-min fairness power control}

Optimal max-min fairness power control is of practical interest and it has been extensively investigated for \Gls{mamimo} systems \cite{2016Marzetta,2017Bjornsonbook}.
%
The \gls{mmf} optimization problem subject to average power constraints at the BS is formulated as
\begin{equation}
    \begin{split}
        \mathcal{P}_1: \hspace{0.4cm} &\underset{p_{g}}{\text{maximize}}\hspace{0.2cm} \underset{g \in \mathcal{G},k \in \mathcal{K}_g}{\text{min}} \hspace{0.3cm} \text{SE}_{gk}\\
         & \hspace{0.5cm} \text{s.t.} \hspace{1cm} \sum_{g \in \mathcal{G}} p_{g} \leq P_{\text{DL}}.\\
   		\end{split}
   		\label{eq:MMF_problem}
\end{equation}
Expression \eqref{eq:SE_gk} can be rewritten as
\begin{equation}
    \begin{split}
        \text{SE}_{gk} = \left(1 - \frac{\tau_p}{\tau_c}\right) \text{log}_2\left(1 +\frac{p_g a_{gk}}{\displaystyle\sum_{g' \in \mathcal{G}} p_{g'} b_{gkg'} + \sigma_k^2}\right),
     \end{split}
    \label{eq:SE_gk2}
\end{equation}
where $a_{gk}=\Big|{\mathbb{E}\left[\boldsymbol{h}^{\text{H}}_{gk}\boldsymbol{w}_g\right]}\Big|^2$, $b_{gkg'} = \mathbb{E}\left[\Big|\boldsymbol{h}^{\text{H}}_{gk}\boldsymbol{w}_{g'}\Big|^2\right]$ for $g'\neq g$ and $b_{gkg} = \mathbb{E}\left[\Big|\boldsymbol{h}^{\text{H}}_{gk}\boldsymbol{w}_{g}\Big|^2\right] - \Big|{\mathbb{E}\left[\boldsymbol{h}^{\text{H}}_{gk}\boldsymbol{w}_g\right]}\Big|^2$. 

Maximizing the minimum SE is equivalent to maximizing the minimum SINR. Therefore, we can rewrite $\mathcal{P}_1$ in epigraph form as
\begin{equation}
    \begin{split}
        \mathcal{P}_2: \hspace{0.2cm} &\underset{p_{g}}{\text{maximize}}\hspace{0.3cm} \Gamma\\
         & \hspace{0.4cm} \text{s.t.} \hspace{0.7cm} \frac{p_g a_{gk}}{\displaystyle\sum_{g' \in \mathcal{G}} p_{g'} b_{gkg'} + \sigma_k^2} \geq \Gamma \hspace{0.3cm} \forall k \in \mathcal{K}_g, g \in \mathcal{G}\\
         &\hspace{1.5cm} \sum_{g \in \mathcal{G}} p_{g} \leq P_{\text{DL}},\\
   		\end{split}
   		\label{eq:MMF_problem2}
\end{equation}
where $\Gamma$ is an auxiliary variable that must satisfy the constraint $\gamma_{gk} \geq \Gamma, \ \forall g \in \mathcal{G},$ and $k \in \mathcal{K}_g$.
$\mathcal{P}_2$ is still non-convex since the \gls{sinr} constraint is neither convex nor concave with respect to $p_g$. To overcome such non-convexity, we use a \gls{sca}. Note that for a fixed value of $\Gamma \geq 0$, the \gls{sinr} constraint in $\mathcal{P}_2$ can also be written in a linear form as \[ p_g a_{gk} \geq \Gamma\left(\displaystyle\sum_{g' \in \mathcal{G}_j} p_{g'} b_{gkg'} + \sigma_k^2\right).\]

If $\Gamma$ is fixed, $\mathcal{P}_2$ is a linear feasibility program, and the optimal solution can be efficiently computed by using interior-point methods, for example, with the toolbox CVX \cite{2014cvx}.
Letting $\Gamma$ vary over an \gls{sinr} search range $\{\Gamma_{min}, \Gamma_{max}\}$, the optimal solution can be efficiently computed by using the bisection method \cite{2004Boyd}, in each step solving the corresponding linear feasibility problem for a fixed value of $\Gamma$.

Algorithm \ref{alg:SCA-MMF} describes the \gls{sca} algorithm providing the power allocation that maximizes the minimum \gls{sinr} among the multicast users. It works for a pre-determined subgrouping configuration (i.e., the formation of the $\mathcal{K}_g$ sets of users).


\begin{algorithm}[!t]
\caption{SCA algorithm for optimal max-min fairness power control with multicast user subgroups}\label{alg:SCA-MMF}
\begin{algorithmic}
  \footnotesize	
    \State {\underline{\textbf{Constant}:} \ \ $P_{\text{DL}}$, \ $\varepsilon$}
    \State {\underline{\textbf{Input}:} \ \ $\{a_{gk}\}$, \ $\{b_{gkg'}\}$}
    \State {\underline{\textbf{Initialization}:}}
    
    \State {$\Gamma_{min} \leftarrow 0$}
    \State {$\Gamma_{max} \leftarrow \underset{g,k}{\text{min}}\left(\frac{P_{\text{DL}}a_{gk}}{\sigma_k^2}\right)$}
    \State {$p^*_g \leftarrow 0, \ \forall g \in \mathcal{G}$}\\
    
    \Do 
        \State {$\Gamma = \frac{\Gamma_{max} + \Gamma_{min}}{2}$}
        \State {Solve (\ref{eq:MMF_problem2})}
        \If {feasible}
            \State {$\Gamma_{min} \leftarrow \Gamma$}
            \State {$p_g^* \leftarrow p_g, \ \forall g \in \mathcal{K}$}
        \Else
            \State {$\Gamma_{max} \leftarrow \Gamma$}
        \EndIf
    \doWhile{$\Gamma_{max} - \Gamma_{min} > \varepsilon$}\\
    \State {\underline{\textbf{Output}:} \ \ $\Gamma_{min}, \ p_g^*$}  
\end{algorithmic}
\end{algorithm}

\subsection{Time/frequency schedule in MaMIMO multicasting}
\label{sec:ssmam}
\Gls{mamimo} multicasting exploits spatial multiplexing through \gls{dl} precoding to handle the intra-cell interference among different subgroups \cite{de2019multiuser}. Alternatively, multicast subgroups can be served over orthogonal time/frequency resources (time/frequency scheduling) \cite{2013Araniti,2018delaFuente}. 



According to \cite{2015Araniti}, the optimal number of time/frequency multicast subgroups that maximizes the \gls{ase} is either one or two in multicast \gls{ofdm} systems. Hence, we compare two setups underlying the proposed user subgrouping strategy: \emph{i)} all the multicast users served in the same time/frequency resource (spatial multiplexing), and \emph{ii)} users served over two orthogonal time-frequency intervals (time-frequency multiplexing). In the latter, a set of subgroups is served in a fraction of the time-frequency resources denoted by $0<\theta<1$, whereas the rest of the subgroups is served in a fraction $1-\theta$ of the resources.
This user scheduling takes place by using the K-means algorithm based on the large-scale fading coefficient $\beta_{gk}$. 

The \gls{ase} for the time/frequency schedule is given by 

\begin{equation}
   \ \text{ASE} = \left(1 - \frac{\tau_p}{\tau_c}\right) \sum_{g \in \mathcal{G}} \hat{\theta}_g K_{g} \text{log}_2\left(1 + \hat{\gamma}_{gk}\right),
\label{eq:ASE}
\end{equation}
where $\hat{\theta}_g$ is equal to either $\theta$ or $1-\theta$ depending on which interval the subgroup $g$ is scheduled, and
\begin{equation} \label{eq:scheduling:SINR_gk}
        \hat{\gamma}_{gk}\! =\! \frac{p_g\Big|{\mathbb{E}\left[\boldsymbol{h}^{\text{H}}_{gk}\boldsymbol{w}_g\right]}\Big|^2}{\displaystyle\sum_{g' \in \mathcal{S}_g}^G p_{g'}\mathbb{E}\left[\Big|{\boldsymbol{h}^{\text{H}}_{gk}\boldsymbol{w}_{g'}}\Big|^2\right]\! -\! p_g\Big|{\mathbb{E}\left[\boldsymbol{h}^{\text{H}}_{gk}\boldsymbol{w}_g\right]}\Big|^2\!\! +\! \sigma_k^2},
\end{equation}
and $\mathcal{S}_g$ denotes the set of the indices of the subgroups scheduled in the same time-frequency interval as subroup $g$, i.e., the subgroups interfering with subgroup $g$.


\section{Numerical results}
\label{sec:results}

In this section, we use numerical simulations to assess the performance of our multicast \gls{mamimo} subgrouping strategy. We have employed different configurations of multicast users, along with a cell of $200$ m. We have placed the users in clusters with a radius of $2$ m to assess users' effect with similar channel characteristics. We have employed spatially correlated fading channels based on each user's nominal angle to the \gls{bs} and an \gls{asd} of $10
^{\circ}$. The path-loss (in dB) is given by $32.4 + 20\text{log}_{10}(f) + 37.6\text{log}_{10}(d)$, where $f$ is the operating frequency in GHz, and $d$ is the 2D distance between the \gls{bs} and the user in meters \cite{2018SadeghiTWC1}. We have modeled a correlated shadowing with a variance of $10$ dB, which presents a low inter-cluster correlation and an extremely high intra-cluster one. We have employed a \gls{ula} with $M$ transmit antennas at the \gls{bs}. The total amount of power available for \gls{dl} transmission is $2$ W, the \gls{ul} pilot power is $1$ W, and the length of the pilot sequence is $G$ (i.e., the number of multicast subgroups). We consider a noise power spectral density of $-174$ dBm/Hz, a receiver noise figure of $7$ dB, and an operating bandwidth of $20$ MHz at a carrier frequency of $2$ GHz (e.g., applicable in sub-6 GHz-\gls{nlos} scenarios). We assumed a channel coherence of $200$ time/frequency samples. We have run Monte Carlo simulations for different configurations of multicast users, over $50$ random spatial distributions with $50$ channel realizations.

\begin{figure}[t]
\centering
\includegraphics[width=\columnwidth]{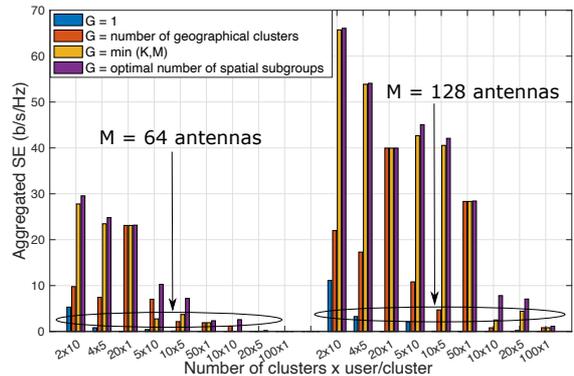}
\caption{Aggregated \gls{se} using spatial subgrouping in only 1 time/freq block. Comparison of using 1 multicast transmission, number of geographical clusters, unicast, or optimum number of multicast subgroups. Different random distributions of users and clusters with 64 and 128 transmit antennas.}
\label{fig:1t_f}
\end{figure}

First, we analyze the subgrouping based on spatial multiplexing without the time/frequency schedule. Fig. \ref{fig:1t_f} presents the \gls{ase} of the multicast \gls{mamimo} service, using the subgrouping strategy for different distributions of users in clusters uniformly and randomly placed along with the cell. We compare the utilization of various criteria to select the number of spatial multicast subgroups (only 1 subgroup, the number of geographical clusters of users, unicast transmissions to every user, and the optimal number of multicast subgroups based on matrix $\boldsymbol{R}_{gk}$). We can observe how the \gls{ase} decreases with the dispersion of the users, i.e., $20$ users placed in $2$ clusters of $10$ users present higher \gls{ase} than $4$ clusters of $5$ users and even more than $20$ uncorrelated users. When the number of users is low compared to the number of transmit antennas, i.e., $20$ users with $64$ transmit antennas, $20$ and $50$ users with $128$ transmit antennas, the \gls{ase} achieved using unicast transmissions is close to the optimal results achieved with our spatial subgrouping proposal. However, when the number of users increases and these users are placed in clusters, the utilization of multicast spatial subgrouping offers notably better performance than unicast transmissions. In any case, the higher the number of users, the lower the \gls{ase}. Consequently, individual \gls{se} is significantly reduced. A large number of \gls{mamimo} spatially multiplexed transmissions and the utilization of a common precoding vector for many multicast users lead to a low channel gain when a high number of multicast users are served using \gls{mamimo} transmissions in the same time/frequency interval.  
 
\begin{figure}[t]
\centering
\includegraphics[width=\columnwidth]{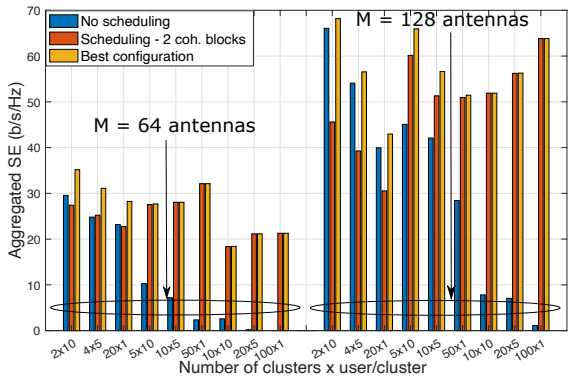}
\caption{Aggregated \gls{se} using spatial subgrouping in only 1 time/freq block, 2 time/freq blocks, or the best configuration. Different random distributions of users and clusters with 64 and 128 transmit antennas.}
\label{fig:1tf_vs_2tf}
\end{figure}

To overcome this limitation, we test the proposal of combining spatial and time/frequency subgrouping. Fig. \ref{fig:1tf_vs_2tf} shows the \gls{ase} for the initial configurations employing only optimal spatial subgrouping and the combination of spatial subgrouping with scheduling using two intervals (using one or two intervals provides higher \gls{ase} depending on the users' distribution). Observing the scenario with a more considerable distance between the number of users and transmit antennas, i.e., $20$ users and $128$ transmit antennas, the option of using only spatial multiplexing is almost always the optimal configuration. Nevertheless, when this relation is diminishing, the design splitting the users suffering from highly different large-scale fading coefficient (${\beta}_{gk}$) into two time/frequency scheduling resources is hugely beneficial. Note that $100$ uncorrelated users cannot be served by $64$ transmit antenna using \gls{mamimo} spatial multiplexing. However, splitting these $100$ users into two scheduling blocks allows the system to deliver the multicast service with an extraordinary improvement in the \gls{ase}.

\begin{figure}[t]
\centering
\includegraphics[width=0.9\columnwidth]{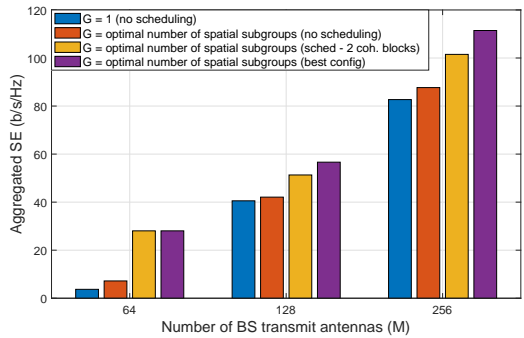}
\caption{\Gls{ase} versus the number of \gls{bs} transmit antennas. Comparison among unicast transmissions and subgrouping with only spatial multiplexing, and with time/frequency scheduling. Config: $10$ clusters x $5$ users/cluster.}
\label{fig:Num_tx}
\end{figure}

Finally, we employ the configuration with $10$ clusters of $5$ users/cluster to evaluate the impact of the number of \gls{bs} antennas on the subgrouping strategy's performance. Fig. \ref{fig:Num_tx} illustrates the \gls{ase} for a variable number of \gls{bs} antennas. We assess the performance of these options: \emph{i}) a unique multicast group ($G=1$) without the time/frequency schedule, \emph{ii}) the optimal number of spatial subgroups without the time/frequency schedule, \emph{iii}) the optimal number of spatial subgroups with the time/frequency schedule, and \emph{iv}) the \emph{best configuration} between \emph{ii}) and \emph{iii}).

When the number of users is comparable with the number of transmit antennas (i.e., $64$ antennas), using two scheduling intervals provides the highest \gls{ase} in almost every users' distribution. Hence, the average performance of scheduling and \emph{best config} options are practically equal. As we increase the number of transmit antennas and keep the users' deployment, spatial multiplexing without scheduling becomes the best configuration in some users' distributions. Thus, the maximum \gls{ase} for some users' allocations is obtained using spatial multiplexing without scheduling and with scheduling for others. Consequently, the larger the transmit antennas, the higher the improvements of using the \emph{best config} option.

\section{Conclusion}
\label{sec:conclusion}

This work studies the multicast transmission in a \gls{mamimo} system, considering spatially correlated fading channels. Subgrouping the multicast users, based on their large-scale spatial correlation matrices, allows the system to improve the channel estimation using common pilots and precoding processes. We have used a \gls{dl} power allocation scheme based on \gls{mmf}. 

We conclude that the optimal subgroup configuration highly depends on random users' distribution and the spatial channel correlation. Correlation matrices information allows the system to create multicast subgroups in scenarios where the users are randomly placed in clusters. Nevertheless, the \gls{ase} dramatically drops when the number of users increases, mostly where they are not set in clusters. Our proposal of splitting the multicast users into two time/frequency schedule blocks based on their large-scale fading coefficient presents a significant improvement in the \gls{ase}. Hence, a resource allocation strategy that first decides using one or two time/frequency schedule blocks and then creates spatial subgroups provides an attractive improvement in the \gls{ase} results.

\section{Acknowledgement}
This work is supported by the “José Castillejo” program, Spanish Ministry of Education and Professional Training. 

\bibliographystyle{IEEEtran}
\bibliography{main}
\end{document}